# An Empirical Study of Bugs in Eclipse Stable Internal Interfaces

Simon Kawuma[1], Evarist Nabaasa[2], David Bamutura Sabiiti[3], Vicent Mabirizi[4]

*Abstract*: **The Eclipse framework is a popular and widely used framework that has been evolving for over a decade. The framework provides both stable interfaces (APIs) and unstable interfaces (non-APIs). Despite being discouraged by Eclipse, application developers often use non-APIs which cause their systems to fail when ported to new framework releases. Previous studies showed that applications using relatively old non-APIs are more likely to be compatible with new releases compared to the ones that used newly introduced non-APIs. Furthermore, from our previous study about the stability of Eclipse internal interfaces, we discovered that there exist 327K stable non-API methods as the Eclipse framework evolves. In the same study, we recommended that 327K stable non-API methods can be used by Eclipse interface providers as possible candidates for promotion to stable interfaces. However, since non-APIs are unsupported and considered to be immature i.e., can contain bugs, to this end there exist a need to first investigate the stable non-APIs for possible bugs before they can be promoted to APIs. In this study, we empirically investigated the stable non-API for possible bugs using Sonarqube software quality tool. We discovered that over 79.8% classes containing old stable non-APIs methods have zero bugs. Results from this study can be used by both interface providers and users as a starting point to analyze which interfaces are well tested and also estimate how much work could be involved when performing bug fixing for a given eclipse release.**

*Keywords*: *Eclipse, Interfaces, Stability, Promotion, Faults, Bugs, Evolution.*

## I. INTRODUCTION

Application developers build their systems on top of frameworks and libraries [1]. Building applications this way fosters reuse of functionality [2] and increases productivity [3]. This is why large application frameworks such as Eclipse [4] MSDN [5], jBPM [6], JUnit [7] commonly provide public (stable) interfaces (APIs) to application developers. In addition to public (stable) interfaces all these frameworks also provide unstable internal interfaces (non-APIs). One widely used and adopted application framework is the Eclipse application framework. Eclipse is a large and complex open-source software system used by thousands of application developers. Eclipse has been evolving for over a decade producing major and minor releases. Eclipse, jBPM, jUnit, all adopt the convention of internal interfaces by using sub-string internal in their package names while JDK non-APIs packages start with the substring sun.

Eclipse Framework developers discourage the use of non-APIs because they may be immature, unsupported, and subject to change or removal without notice [4], [5], [7], [8]. Supporting these recommendations, previous empirical studies have shown that when the Eclipse application framework evolves, APIs do not cause compatibility failures in applications that solely depend on them [9], while non-APIs cause compatibility failures in applications that depend on them [9], [10]. Despite non-API being discouraged and causing compatibility failures, usage of non-APIs is common. Businge et al. observed that about 44% of 512 Eclipse plug-ins use non-APIs [11], [12]. Furthermore, application developers claim that they cannot find APIs with the functionality they require among APIs and therefore feel compelled to use non-APIs [13]. Much as the developers discourage the use of non-APIs, they do know that application developers use them. In our preliminary study about the non-APIs, we observed twice as many fully qualified non-API methods compared to APIs methods [14]. As a solution to mitigate discussed risks and help client developers, API producers may promote some internal interfaces to public ones. However, Hora et al. [15] discovered that promotion occurs slowly causing a delay to client developers to benefit from stable and supported interfaces. The authors further state that slow promotion results from API producers having no assistance in identifying public interface candidates (i.e., internal interfaces that should be public). Kawuma et.al [16] further confirmed that indeed the pace at which non-APIs are promoted to APIs is slow and promotion take long.

In our recent study [17], using clone detection techniques we investigated the stability of Eclipse internal interfaces over subsequent Eclipse releases. In the same study, we provided a dataset of 327K old stable non-API methods that can be used by framework developer as possible candidates for promotion. However, since non-APIs are considered to be immature i.e., can contain bugs, there exist a need to investigate the stable non-API for possible bugs before they can be considered for promotion. Using SonarQube software quality tool [18], this study discovered that there exist bug-free stable non-APIs among the stable non-API discovered in our previous study [17]. In addition to their stability and being bug-free these stable non-APIs can be good candidates for promotion to stable APIs and should be recommended to application developers.

* Correspondence Author
**Simon Kawuma\***, Software Engineering department, Mbarara University of Science and Technology, Mbarara, Uganda. Email: simon.kawuma@must.ac.ug
**Evarist Nabaasa**, Computer Science department, Mbarara University of Science and Technology, Mbarara, Uganda. Email: enabaasa@must.ac.ug
**David Bamutura Sabiiti**, Computer Science department, Mbarara University of Science and Technology, Mbarara, Uganda. Email: dbamutura@must.ac.ug
**Vincent Mabirizi**, Department of Information Technology and Computer Science, Kabale University, Kabale, Uganda. Email: mvicent151@gmail.com

The remainder of this paper is organized as follows: Section II presents the background information on Eclipse interfaces, bugs and research objective and question. Section III discusses the experimental setup of our study. Section IV discusses the results and findings of our study. Section V presents threats to the validity of our study, while Section VI provides an overview of the related work. Finally, Section VII concludes the paper and outlines some avenues for future work.

## II. Background

**Eclipse non-APIs** are internal implementation artifacts that according to Eclipse naming convention [4] are found in packages with the substring internal in the fully qualified name. These internal implementation artifacts include public Java classes or interfaces, or public, protected methods, or fields in such a class or interface. Usage of non-APIs is strongly discouraged since they may be unstable [19]. Eclipse clearly states that clients who think they must use these non-APIs do it at their own risk as non-APIs are subject to arbitrary change or removal without notice. Eclipse does not usually provide documentation and support to these non-APIs.

**Eclipse APIs** are public Java classes or interfaces that can be found in packages that do not contain the segment internal in the fully qualified package name, a public or protected method, or field in such a class or interface. Eclipse states that, the APIs are considered to be stable and therefore can be used by any application developer without any risk. Furthermore, Eclipse also provides documentation and support for these APIs.

**Error** are the most basic discrepancies found by the team of testers. These are the mistakes made by the software developer or programmer, while preparing the code or design of the software. Errors are mainly a deviation from the results expected by the team, which further changes the functionality of the software [20]. Error can also be defined as incorrect or missing human action that result in system/component containing a fault. Examples include omission or misinterpretation of user requirements in a product specification, incorrect translation, or omission of a requirement in the design specification, Miscalculation of some values, Mistakes in design or requirement activities, Discrepancy between actual and expected results.

**Fault** is Introduced in the software because of an error, fault is another discrepancy found by the team of testers during the process of software testing. Unlike error, the reason for a fault to occur while implementing the software is not because of a miscalculation of some value or discrepancy between actual and expected result, but is merely a manifestation of an error in a software. Moreover, a fault in the software system inhibits it from performing its intended function and forces the system to act in an unanticipated manner [20]. Faults in a system can be raised because of Discrepancy or issue in the code that causes the failure of the system/program, introduction of an incorrect step, process, or data definition. Fault can also be an anomaly or irregularity in the software, which makes the software behave in an incorrect way and not as per the stated requirements.

**Bugs** are the most integral part of a software system and can be termed as the errors, flaws, and faults present in the computer program that impact the performance as well as the functionality of the software can cause it to deliver incorrect and unexpected results. These not only impact the performance of the software, but also cause it to behave in an unanticipated way [20].

**SonarQube:** In this research study, we used SonarQube which is one of the most common Open-Source static code analysis tools adopted both in academia [21], [22] and in industry [23]. SonarQube is provided as a service from the sonar-cloud.io platform or it can be downloaded and executed on a private server. SonarQube calculates several metrics such as the number of lines of code and the code complexity, and verifies the code's compliance against a specific set of coding rules defined for most common development languages [24]. In case the analyzed source code violates a coding rule or if a metric is outside a predefined threshold, SonarQube generates an issue. SonarQube includes Reliability, Maintainability and Security rules. SonarQube has separate sets of rules for the most common development languages such as Java, Python, C++, and JavaScript. In this research, we used SonarQube version 8.2 which has more than 500 rules for Java. The complete list of rules is available online[1]. Reliability rules, also named bugs create issues (code violations) that represents something wrong in the code and that will soon be reflected in a bug. The severity of the bugs can be categorized by their possible impact, either on the system or on the developer's productivity. SonarQube categorizes the identified bugs into five types namely blocker, critical, major and minor [18] as explained below;

- **Blocker:** A bug of this kind might make the whole application unstable in production. For example, calling garbage collector, not closing a socket, memory leak, unclosed JDBC connection etc.
- **Critical:** A bug of this kind might lead to an unexpected behavior in production without impacting the integrity of the whole application. For example, Null-PointerException, badly caught exceptions, division by zero as a denominator etc.
- **Major:** A bug of this kind might have a substantial impact on productivity. For example, Null pointers should not be dereferenced, too complex methods, package cycles.
- **Minor:** A bug of this kind might have a potential but minor impact on productivity. For example, finalizer does nothing but call superclass finalizer, lines should not be too long, switch statements should have at least 3 cases.

Blocker and critical bugs might impact negatively the system, with blocker bugs having a higher probability compared to critical ones. SonarQube also recommends immediately reviewing blocker and critical issues. Major bugs can highly impact the productivity of a developer, while minor ones have little impact.

---

[1] https://rules.sonarsource.com/java

### III. STUDY DESIGN AND EXPERIMENT SETUP

This section presents the experimental setup of how the data for the research questions was collected. To allow independent replication and verification of the study, we provide a full replication package online and *the detailed lists of classes containing bug-free stable non-API methods in different Eclipse new releases is available on-line*[2].

#### A. Goals and Research Question

This study is an extension of our recent work where we studied the stability of internal interfaces [17] during the evolution of the Eclipse framework. In this study, we analyzed stable internal interfaces for possible bugs. The study aimed at recommending the bug-free stable internal interface to Framework developers for promotion to stable APIs. We understand that Eclipse providers normally perform non-API promotion at the level of a class. Promotion at the method level seems to be fine grained thus from our previous study [17], we summarized the stable non-API methods that are possible candidates for promotion to APIs to the class-level. For example, instead of promoting a single non-API method, the providers can promote groups of non-API methods that are stable together in a single class. With this in mind, we formulated the following research question to guide this study as follows: **RQ: Can we find bug-free stable internal Interfaces in Eclipse Frameworks?** Eclipse being large and complex, it is possible that there exist stable internal interfaces which can be recommended to developers. Indeed, in this research, we discovered that over 79.8% of the classes containing the stable internal interfaces (non-APIs) methods have zero bugs and thus these would be good candidate for promotion to stable APIs in all studied Eclipse release.

**Contribution of our work:** findings from this study are interesting because users of the framework are not aware of bug-free stable non-APIs and they would fear to use the stable non-API methods while developing their applications because they are still tagged as unstable interfaces. Findings from this study lays a foundation for API providers to promote the bug-free stable non-APIs to APIs. Furthermore, we provide a public dataset[3] of bug-free stable non-API classes that can be used by both interface providers and users. We recommend that *Interface providers and users* can use our dataset findings as starting point to choose bug-free interfaces to use in their application instead of randomly using interfaces which might have bugs.

#### B. Eclipse Release Collection

In this section, we explain the data sources of our study. Our study is based on 16 Eclipse SDK major releases from Eclipse project Archive website [25], [26] until Eclipse 4.6.

Table I presents the Eclipse major releases with their corresponding release dates. This research study considered Eclipse as a subject of study because it is widely used and adopted open-source framework and thus it will continue attracting more developers. The Eclipse framework is constantly evolving with new version released every after 3 months. This creates an opportunity to study bug evolutionary trends as the framework evolves. This research focused on Eclipse major versions because as the framework evolves from one major version to another, new project, sub-projects, packages, classes, interfaces, fields and methods are added, changed or deleted from the framework.

| Major Releases | Release Date | Major Release | Release Date |
|---|---|---|---|
| E-1.0 | 07-Nov-01 | E-3.7 | 13-Jun-11 |
| E-2.0 | 27-Jun-02 | E-4.2 | 08-Jun-12 |
| E-2.1 | 27-Mar-03 | E-4.3 | 05-Jun-13 |
| E-3.0 | 25-Jun-04 | E-4.4 | 06-Jun-14 |
| E-3.1 | 27-Jun-05 | E-4.5 | 03-Jun-15 |
| E-3.2 | 29-Jun-06 | E-4.6 | 06-Jun-16 |
| E-3.3 | 25-Jun-07 | | |
| E-3.4 | 17-Jun-08 | | |
| E-3.5 | 11-Jun-09 | | |
| E-3.6 | 08-Jun-10 | | |

Table I: Eclipse major releases and their corresponding release dates

#### C. Data Collection for Classes containing stable non-APIs

| | E-4.2 | E-4.3 | E-4.4 | E-4.5 | E-4.6 |
|---|---|---|---|---|---|
| Stable non-APIs | TC | TC | TC | TC | TC |
| E-1.0 | 170 | 169 | 163 | 160 | 149 |
| E-2.0 | 299 | 293 | 287 | 279 | 267 |
| E-2.1 | 187 | 184 | 173 | 166 | 155 |
| E-3.0 | 860 | 840 | 788 | 751 | 696 |
| E-3.1 | 655 | 632 | 583 | 564 | 540 |
| E-3.2 | 894 | 875 | 837 | 805 | 753 |
| E-3.3 | 690 | 658 | 625 | 596 | 544 |
| E-3.4 | 642 | 627 | 593 | 556 | 517 |
| E-3.5 | 734 | 716 | 593 | 664 | 622 |
| E-3.6 | 431 | 424 | 394 | 378 | 351 |

Table II: Total number stable non-API classes of old Eclipse releases (i.e., E-1.0 to E-3.6) in the new Eclipse releases (i.e. E-4.2 to E-4.6)

Table II present the total number of classes of old Eclipse releases containing stable non-API methods whose bug is under investigation. The first row of Table II shows the different new Eclipse releases $E_{new}$ i.e E-4.6 to E-4.4 which have stable non-API classes of the old Eclipse releases $E_{old}$ in the first column. Column labeled TC (Total Classes) shows the total number of stable non-API classes of old Eclipse releases $E_{old}$ in the new Eclipse releases $E_{new}$. For example, looking at row E-1.0 and column E-4.2, the value in cell (row–E-1.0, column–TC) =170, indicates that there exist 170 stable non-API classes in Eclipse E-4.2 that originated from Eclipse E-1.0. Data in Table II was extracted and obtained as part of our recent work on the stability of non-API methods [17].

---

[2] https://sites.google.com/must.ac.ug/dataset-fault-free-stable-non-/home
[3] https://sites.google.com/must.ac.ug/dataset-fault-free-stable-non-/home

### D. *Data Collection and Extraction of Bugs*

In this section, we present how we extracted data for research question **RQ**. We used SonarQube tool (version-8.2) [18] to extract information about bugs in the different Eclipse releases. We rely on this tool because it is broadly used by thousands of users in academic research settings [21], [22], [27] and in industry [23]. We configured and ran SonarQube on a local computer and used its web interface to monitors the analysis results. We considered 125 reliability rules provided in sonarQube version 8.2 to detect bugs. when any of the rules is violated, then that particular source code manifest as a bug. We investigated the total number of bugs in the new Eclipse releases which have the identified stable non-APIs.

| eclipse-1.0 stable bad interfaces in eclipse-4.6 | FRR |
|---|---|
| eclipse-4.6/org/eclipse/core/internal/resources/Synchronizer.java | E |
| eclipse-4.6/org/eclipse/team/internal/ui/Utils.java | D |
| eclipse-4.6/org/eclipse/ui/internal/ide/dialogs/WelcomeItem.java | C |
| eclipse-4.6/org/eclipse/team/internal/core/StringMatcher.java | B |
| eclipse-4.6/org/eclipse/pde/internal/swt/tools/IconExe.java | E |
| eclipse-4.6/org/eclipse/core/internal/dtree/DataDeltaNode.java | A |
| eclipse-4.6/org/eclipse/compare/internal/TabFolderLayout.java | A |

Table III: SonarQube Sample Report

SonarQube tool takes as input a source file containing classes to detect possible bugs and reported specific points in the class where bugs are. The tool produces an output report as shown in Table III. Each file in the report has a File Reliability Rating (FRR) assigned by SonarQube tool depending on the nature and number of bugs found in the class of source files under investigation. For example, from Table III, the last two rows have files with File Reliability Rating (FRR) of A i.e., they have no bugs. The tool counts the number of bugs reported in each class. Additionally, the tool further rates the file reliability rating of the class as follows: A–Zero bug, B– at least one minor bug, C–at least one major bug, D–at least one critical bug and E–at least one blocker bug. In this study, we considered the total number of bugs reported in each class for each Eclipse release. We then focused on the class that had zero bugs i.e., with rating A. To determine the percentage of bug-free stable non-API classes, we express the number of classes with rating A as a fraction of the total number of stable non-API classes for a given Eclipse old release in the new Eclipse release.

### IV. RESULT AND DISCUSSION

In this section we present the results and analysis of the extracted data in Section III-A to address the research question. Figures 1 presents results corresponding to percentage of bug-free stable non-API classes of the old Eclipse releases (E-1.0 to E-3.6) in the new Eclipse releases (E-4.2 to E-4.6). In figure 1, for each new Eclipse releases there are ten bars, and each bar presents the percentage of stable non-APIs classes of the old Eclipse releases (E-1.0 to E-3.6) in the new Eclipse releases. Focusing on Eclipse-4.6 (E-4.6), the first bar presents the percentage of bug-free classes containing stable non-APIs of Eclipse-1.0 (E-1.0) in new Eclipse-4.6 (E-4.6). For example, 86.6% of the total classes of E-1.0 in E-4.6 have no bugs. Looking at the bars in figure 1 we observe that the percentage of classes containing stable non-APIs methods with zero bug ranges from 79.8% to 90.6% for all the studied Eclipse new releases i.e. (E-4.2 to E-4.6). Since non-APIs are considered to be immature, unsupported and subject to change and even can be deleted from the framework [4], [5], [7], [8], [28], one would expect to see number of bugs in the stable non-API classes. However, from our investigation, we have discovered that over 79.76% of classes containing stable non-APIs methods have zero bugs for all the studied Eclipse releases.

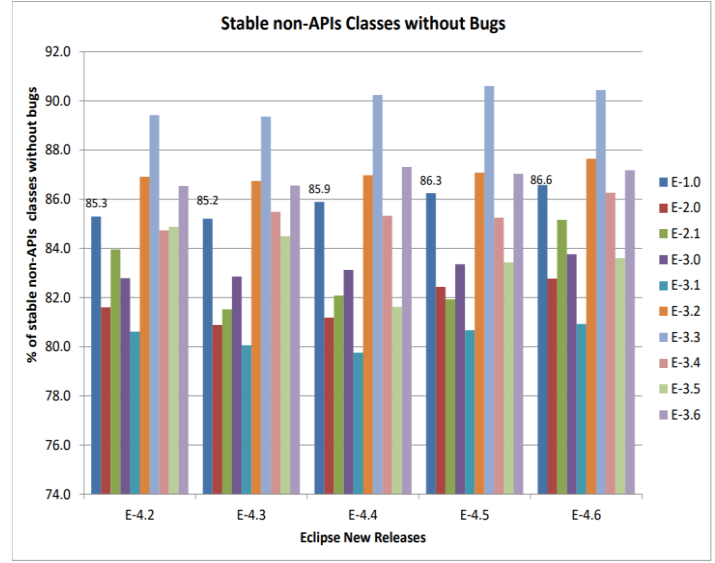

Figure 1: Percentage of stable non-API Classes without bugs

### V. THREATS TO VALIDITY

As any other empirical study, our analysis may have been affected by validity threats. We categorize the possible threats into construct, internal and external validity.

**Construct validity** focuses on how accurately the metrics utilized measure the phenomena of interest. The methodology used to measure the percentage of bug-free interfaces in Eclipse is subjected to construct validity. The reason being that we only use classes in our computations yet there are other objects we ignore, for example, methods, variable declarations etc.

**Internal validity** threat related to the tool (i.e., SonarQube) used to extract the data used in our experiments. It is possible that results could differ if a different tool was used. Like any other static analysis tool, the SonarQube tool we used does not have a 100% precision.

**External Validity** is related to the possibility to generalize our results. We focused on the analysis of widely adopted and large-scale framework. Therefore, Eclipse SDK Framework is a credible and representative case study. The framework is open source and thus its source code is easily accessible. Despite these observations, our findings as usual in empirical software engineering study cannot be directly generalized to other systems, specifically to systems implemented in other programming languages other than java.

## VI. RELATED WORK

In this section we discussed how the current work relates to the previous work. The previous studies by Businge et al. [9],[13], [29] were based on empirical analysis of the co-evolution of the Eclipse SDK framework and its third-party plug-ins (ETPs). During the evolution of the framework, the authors studied how the changes in the Eclipse interfaces used by the ETPs, affect compatibility of the ETPs in forthcoming framework releases. The authors only used open-source ETPs in the study and the analysis was based on the source code. One of previous studies by Businge et al. [13] was based on analysis of a survey, where they complement other previous studies by including commercial ETPs and taking into account human aspects. One of the major findings of the previous studies was that interface users are continuously using unstable interfaces and the reason for using these unstable interfaces was because there exist no alternative stable interfaces offering the same functionality. Indeed, Kawuma et al. showed that less than 1% APIs offer the same or similar functionality as non-APIs [14]. The earlier studies by Businge et. al studied both APIs and non-APIs interfaces but they did not look at finding bugs in internal interface(non-APIs) as compared to our study.

From our recent study Businge et.al [17], using a clone detection tool, we looked at the stability of internal interface as the Eclipse framework evolves. We discovered that 327K stable internal Interfaces and we recommended them as possible candidate for promotion. Another study that is directly related to [17] is that of Hora et al. [15], in this study the authors investigated the transition from internal to public interfaces. They carried out their investigation on Eclipse (JDT), JUnit, and Hibernate. Their main aim was to study the transition from internal to public interfaces (i.e., internal interface promotion). They detect internal interface promotion when these two conditions are satisfied: 1) there is at least a file change that removes only one reference to Internal and adds only one reference to Public, and 2) the class names of the references remain the same or have a suffix/prefix added/removed. They discovered that 7% of 2,277 of internal interfaces are promoted to public interfaces. They also found that the promoted interfaces have more clients. They also predicted internal interface promotion with precision between 50%–80%, recall 26%–82%, and AUC 74%–85%. Finally, by applying their predictor on the last version of the analyzed systems, they automatically detected 382 public interface candidates. Our study and this study both aim at identifying internal interfaces that are candidates of promotion. A similar study by Kawuma et. [16] discovered that indeed the pace at which non-APIs are promoted to APIs is slow and promotion take long. Although studies in [17], [15] and [16] identified and recommended internal Interface for promotion, none of the above authors studied existence bugs in stable internal interfaces.

Businge et al. [30] studied the relationship between code authorship and fault-proneness of android applications, they investigated whether android applications with few major contributors have more or less faults compared to application with larger number of developers that do minor contributions. They discovered that android applications with higher level of code authorship among contributors experience fewer faults. Bird et al. [31] found that a module that is written by many minor authors is more likely to have faults in future. Latifa et al. [32] carried out a study on projects; ANT, ArgoUML and Hibernate to establish the relationship between lexical smell and software quality as well as their interaction with with design smells. They discovered 29 smells out of which 13 were design smell and 16 were lexical smells. in addition, they found out that lexical smells can make clases with design smell more fault-prone and that classes containing design smell only are more fault-prone than classes with lexical smell only. Bavota et al. [33] conducted a study on 5,848 free android app to investigate how the apps user ratings correlate with the fault- and change proneness of the APIs such app relied on. From their study, they discovered that apps having high user ratings use APIs that are less fault- and change-prone than the apps used by low rated apps. Assaduzzaman et al. [34] mined changes and bug reports in Android to identify changes that introduced the bugs. The links between bugs and changes were identied by looking for keywords in commit messages, and by comparing the textual similarity between the reports and the commit messages. None of the above studies focused on internal specifically looking at existence of bugs.

API evolution has also been studied for many other platforms. Jezek et al. [35] investigated the API changes and their impacts on Java programs. They found out that API instability is common and will eventually cause problems. Hora et al. [36] studied how developers react to API evolution for the Pharo system, they discovered that API evolution can have a large impact on a software ecosystem in terms of client systems, methods, and developers. Hou and Yao [37] explored the intent behind API evolution by by analyzing the evolution of a production API (AWT/Swing) in detail. The authors discovered that a large part of API evolution is minor correctives, for example, fixing naming problems (spelling errors, weak names) or issues related to violations of general design principles such as coupling and encapsulation.

## VII. CONCLUSION

In this study we have carried out an investigation on the stable Eclipse non-APIs. We have observed that indeed majority of classes containing the stable non-APIs methods have no bugs. We recommend that *Interface providers* can use our findings as a starting point to promote bug-free stable non-APIs to APIs. This will increase the number of stable interfaces which can be used by application developers. In a follow up study, we are planning to study the popularity of the identified bug-free interfaces by looking at both their internal and external usage. Internal interface usage can be determined by looking at how many packages and libraries in Eclipse framework use the identified bug-free interfaces. External usage can be determined by looking at how many applications on Github use bug-free interfaces. Similarly, external usage can be measured by looking at the number of developers who have used or touched a particular bug-free interfaces. Although majority of the interfaces have no bugs, this does not qualify them meet all the other software quality metrics we plan to carry out an investigation and ascertain the other software quality of the identified bug-free interfaces by looking at more parameters like technical debt, complexity, documentation and maintainability.

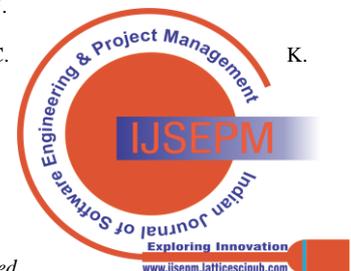

## AUTHORS PROFILE

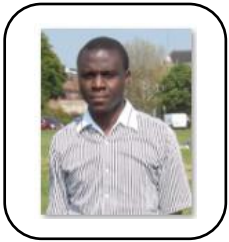

Simon Kawuma received his PhD in Software Engineering from Mbarara University of Science and Technology, MSc degree in Embedded Computing System at Norwegian University of Science and Technology Norway and University of Southampton United Kingdom in and BSc. Computer Science from Mbarara University. His research interests include Empirical software engineering, software maintenance, Evolution, Mining Software Ecosystems Software Quality, modelling and Simulation and.

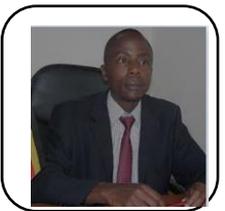

Evarist Nabaasa is a senior lecturer and Dean Faculty of Computing and Informatics at Mbarara University of Science and Technology. He earned his Ph.D. in Computer Science from Mbarara University of Science and Technology in 2015. His research interests include software engineering, Computer Networks, and algorithms.

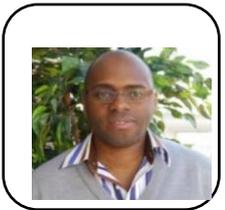

David Bamutura is a lecturer in the Department of Computer Science Department at the Institute of Computer Science, Mbarara University of Science and Technology since February, 2014. Previously, I was an Assistant Lecturer in the Department of Networks at the College of Computing and Information Sciences, Makerere University for two years (2011-2013). My research interests lie in several areas such as Natural Language Processing (NLP), Computer Systems Architecture and Programming languages, Software Engineering.

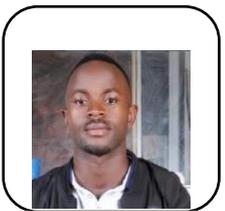

Vicent Mabirizi an Assistant lecturer in the Department of Information Technology and Computer Science, Kabale University. He holds a Master of Science in Health Information Technology and Bachelor of Information Technology from Mbarara University of Science and Technology. He has interest in Software Engineering and AI and Data science.